# The Environments of $z < 0.3$ QSOs

B.J. Boyle [1], R.J.Smith[2], S.J. Maddox[1]
[1] *Royal Greenwich Observatory, Cambridge, UK.*
[2] *Institute of Astronomy, Cambridge, UK.*

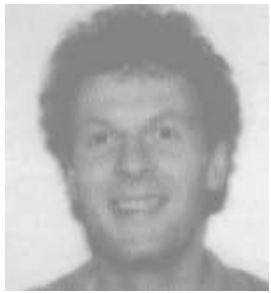

**Abstract**

Based on a cross-correlation analysis of X-ray selected QSOs with faint ($B_J < 20.5$) galaxies, we find a $5\sigma$ galaxy excess around low-redshift ($z < 0.3$) QSOs with an amplitude identical to that of the galaxy angular correlation function. The similarity between QSO-galaxy clustering and galaxy-galaxy clustering suggests that QSOs are unbiased with respect to galaxies and make useful tracers of large-scale structure in the Universe.

## 1 Introduction

The clustering of QSOs can, in principle, place useful constraints on the form and evolution of structure in the Universe [11], particularly at redshifts ($z > 0.3$) and scales ($> 100h^{-1}\,{\rm Mpc}$) which can not directly be probed with galaxies. However, existing studies of QSO clustering are limited to relatively small samples of QSOs ([1], [7], [11]) typically containing less than 1500 objects. Although most of these studies agree that QSOs exhibit significant clustering at small scales ($< 20h^{-1}\,{\rm Mpc}$) with a mean correlation length $r_0 \sim 6h^{-1}\,{\rm Mpc}$, there is little concencus over the evolution with redshift in the strength of this clustering or in the strength of clustering at larger scales ($> 20h^{-1}\,{\rm Mpc}$). This is primarily due to the lack of observational data. Fortunately, with the imminent availability of wide-field multi-object spectrographs such as the 2dF [14] and Sloane Digitial Sky Survey [5], significantly larger QSO surveys should be forthcoming within the next 2-3 years. For example, there is a current proposal to conduct a QSO survey on the AAT 2dF (P.I.s Boyle and Shanks) which will yield spectra for 30000 $B_J < 21$ UVX-selected QSOs in 40 nights of telescope time. With such a survey it will be possible to measure *directly* the evolution of clustering with redshift, thereby placing limits on galaxy formation models and the value of $\Omega_0$. The survey will also provide an accurate estimate of the strength of QSO clustering at large scales ($> 20h^{-1}\,{\rm Mpc}$), yielding important information on the index of the primordial fluctuation power spectrum, $n$, as well as providing a potential measure of $q_0$ and $\Lambda_0$ [11].

However, QSOs can only be used as reliable probes of large-scale structure if it is understood how they are distributed with respect to galaxies. Existing deep CCD surveys ([2], [4]) indicate that radio-quiet QSOs (which comprise over 95% of the QSO population) inhabit normal galaxy environments over a wide range in redshifts ($0.3 < z < 1.5$). This is in marked contrast to radio-loud QSOs, whose environments undergo strong evolution over the range $0.3 < z < 0.6$ [4] and appear to inhabit rich clusters at higher redshifts [8]. Unfortunately, results on QSO environments at these redshifts are drawn primarily from deep CCD imaging studies and are thus highly dependent on the correction applied to convert *observed* galaxy excess on the sky to *physical* galaxy clustering at the redshift of the QSO. For all but the lowest redshifts ($z \sim 0.3$), this involves an uncertain extrapolation of the galaxy luminosity function and its evolution with redshift. To obtain a more accurate estimate of the environments of QSOs, we have therefore carried out out an investigation of the environments of the lowest redshift ($z < 0.3$) QSOs, using QSO and galaxy samples which have recently become availiable. Full details of this work are to be published in [12].

## 2 Data

We cross-correlated the positions of 169 X-ray selected $z < 0.3$ QSOs identified in the *Einstein Medium Sensitivity Survey* (EMSS, [13]) with the positions of $B_J < 20.5$ galaxies detected from the APM galaxy survey [10] and from the APM Sky Catalogues [6]. The $z < 0.3$ QSOs are predominantly (> 96%) radio-quiet [3] and lie in the absolute magnitude range $-24 < M_B < -21$. From these data-sets we derived the QSO-galaxy cross-correlation function $w_{qg}(\theta)$ as follows:

$$w_{qg}(\theta) = \frac{N_{QG}}{N_{QR}} - 1, \qquad (1)$$

where $N_{QG}$ is the number of galaxies found at a given separation ($\theta$) and $N_{QR}$ is the number of random points expected in the same annulus. Fortuitously, the redshift distribution of the QSO sample is identical to that of $B_J < 20.5$ galaxies, and thus (since both samples have identical projection effects) the amplitude of $w_{qg}(\theta)$ can be directly compared to that of the $B_J < 20.5$ galaxy angular correlation function, $w_{gg}(\theta)$, in order to estimate the relative strength of QSO and galaxy environments.

## 3 Results

In figure 1 we plot the pair-weighted $w_{qg}(\theta)$ obtained from all $z < 0.3$ QSOs in the sample. The positive detection of clustering at small scales ($< 4\,\mathrm{arcmin}$) is significant at the $4.7\sigma$ level (728 independent QSO-galaxy pairs found, 610 pairs expected). Errors bars in figure 1 are based on the rms spread in $w_{qg}(\theta)$ over 5 independent sub-samples. Based on $\chi^2$ fits, the ratio between $w_{qg}(\theta)$ and $w_{gg}(\theta)$ is given by:

$$w_{qg}(\theta)/w_{gg}(\theta) = 1.0^{+0.7}_{-0.4} \qquad (2)$$

where the errors represent the 95% confidence interval. We find no signficant detection of galaxy clustering around a 'control' sample of 154 $z > 0.3$ QSOs selected from the EMSS, indicating that the observed correlation is not an artefact of the data-sets used. Splitting the QSO sample into Northern and Southern fields (for which different galaxy catalogues were used) or into low

($M_B > -22$) and high ($M_B < -22$) luminosity ranges also revealed no significant difference between either sets of sub-samples.

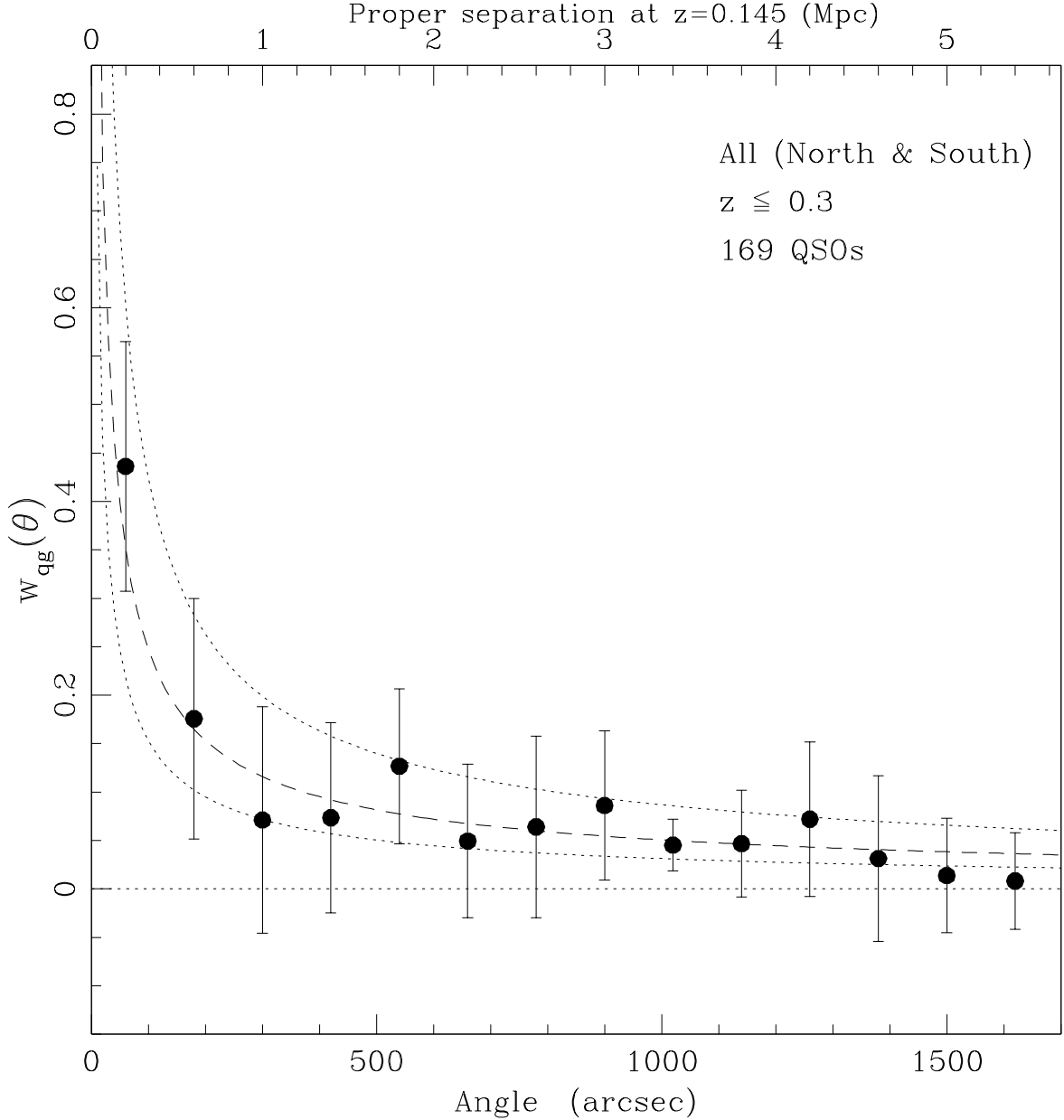

**Figure 1** The QSO-galaxy angular cross-correlation function, $w_{qg}(\theta)$, for 169 $z \leq 0.3$ EMSS QSOs and the APM $B_J \leq 20.5$ galaxies. The dashed curve is the angular $B_J < 20.5$ mag galaxy auto-correlation function, $w_{gg}(\theta)$, derived from the APM galaxy survey [9]. The dotted lines indicate the $\pm 2\sigma$ limits for the fit of $w_{gg}(\theta)$ to $w_{qg}(\theta)$.

## 4  Discussion

From the observed agreement between the amplitudes of $w_{qg}(\theta)$ and $w_{gg}(\theta)$ we conclude that low luminosity ($M_B > -24$) QSOs randomly sample the galaxy distribution at low redshifts ($z < 0.3$). Coupled with previous results ([2], [4]) this implies that radio-quiet QSOs undergo no appreciable change in their environment over a wide range of redshifts ($0 < z < 1.5$) or luminosities ($-27 < M_B < -22$). This suggests that radio-quiet QSOs are as unbiased a tracer of mass in the Universe as galaxies, but have the further advantages of sparse sampling and high luminosity. The agreement between the QSO-galaxy and galaxy-galaxy correlation functions also implies that both QSOs and galaxies are likely to have similar correlation lengths at the present epoch. This is consistent with the observed comoving correlation length of QSOs at $z \sim 1.5$ (which is also the same as that of present day galaxies) only if the QSO correlation length is constant in comoving separation [11]. Although care must be taken in interpreting this result (the comoving separations over which the correlation lengths are derived – $\leq 2$ Mpc for the low redshift QSO-galaxy correlation function and $\leq 10$ Mpc for the high redshift QSO-QSO correlation function – could correspond to different non-linear and linear evolutionary regimes), it is clear that the study of QSO environments at low/intermediate redshifts provides a useful additional constraint on the evolution of structure in the Universe.

**Acknowledgements.** RJS acknowledges the receipt of a PPARC studentship.